\documentclass{article}
\usepackage{spconf,amsmath,graphicx,booktabs,multirow,multicol,cleveref,tabularx,url}

\crefname{section}{§}{§§}
\Crefname{section}{§}{§§}

\title{An Experimental Study: Assessing the Combined Framework of WavLM and BEST-RQ for Text-to-Speech Synthesis}

\nameone{Nielson, V.$^\star$\thanks{$^\star{}$vnielson@perimeterinstitute.ca}}
\nametwo{Hillis, S.}
\addressone{Deepgram Inc., 207 E Washington St, Ann Arbor, Michigan 48104, USA}
\addresstwo{Perimeter Institute for Theoretical Physics (PI), 31 Caroline St N, Waterloo, ON N2L 2Y5, Canada}

\begin{document}
%
\maketitle
\begin{abstract}
We propose a new model architecture specifically suited for text-to-speech (TTS) models. We combine WavLM, a pre-trained self-supervised learning (SSL) speech model, and the BEST-RQ vector quantization framework. We assess the extent to which the more task-agnostic WavLM, coupled with the superior suitability of the simplistic BEST-RQ framework for a wider array of downstream tasks, yields favorable outcomes. Experiments on the LibriSpeech dataset with SUPERB benchmarking assert that the proposed model significantly underperforms. We speculate the underlying reason for this performance is related to the difference between featurizing raw audio waveforms and spectrograms with a quantizer. We discuss the limitations of this approach to better guide future advancements in TTS. 
\end{abstract}
\section{Introduction}
\label{sec:intro}

The substantial progress made in the natural language processing (NLP) domain over the last few years has allowed AI to step into a position of ubiquity and permanence in the world. Whereas before it was the prevailing belief that AI could only service quantitative, cerebral tasks, now it has been made abundantly clear that it is capable of creative tasks as well in a way that resembles human cognitive function. This has been made possible largely by the increased use of self-supervised learning (SSL) techniques. SSL has led to many state-of-the-art NLP systems, not directly due to their time-efficiency and reduction of cost, but because of the scalability that they lead to. Specifically, the need for a scalable solution is dire in NLP systems as, if the goal is to reflect the human neural structure, the amount and breadth of the input data must equal that accessible to a brain in the natural world. Regardless of the amount of effort one invests in enhancing the model or the extensive manual work and time dedicated throughout the process, the capabilities of the model will never surpass the quality level of its underlying input data.  

A significant application of SSL in the field of NLP has recently emerged in Automatic Speech Recognition (ASR) technology. SSL is responsible for the notable strides in terms of accuracy and performance that ASR has made; state-of-the-art ASR systems have achieved remarkable improvements in transcription quality, especially in challenging conditions such as noisy environments, accented speech, and cross-talk. The availability of vast amounts of labeled data and increased computational resources have enabled ASR models to be trained on massive datasets. There has been a strong focus on ASR due to the crucial role of speech-to-text (STT) synthesis in numerous applications, including transcription, dictation, closed caption, or developing tools for inspecting and discovering latent information in the speech (intent classification, emotion recognition, keyword spotting, topic detection, summarization, etc.). However, in the endeavor to explore the entire realm of NLP, focus must also be diverted to text-to-speech (TTS) systems. 

It is notoriously difficult to find TTS systems that are scalable, natural sounding, and fast, let alone to the level of ASR systems. Current systems struggle to achieve even just one of the aforementioned qualities, with prosody being the hardest and rarest achievement. There have been some attempts made, one of the notable ones being the VALL-E (X) neural codec language model cross-lingual developed by Microsoft \cite{zhang2023speak, wang2023neural}. VALL-E (X) is trained on a dataset that includes both text and corresponding audio recordings. In traditional TTS methods, the goal is to directly regress the continuous audio signal from the input text. However, VALL-E (X) takes a different approach; it treats TTS as a conditional language modeling task, where the model learns to generate the most probable sequence of discrete codes conditioned on the input text. One of the key advantages of VALL-E (X) is its ability to learn in context. By leveraging the training data and the discrete codes derived from the audio, VALL-E (X) can capture contextual information and generate more natural and coherent speech. The results from \cite{zhang2023speak} show that only one speech utterance is needed to generate speech that closely matches the unique characteristics of the speaker. More recently, SPEAR-TTS \cite{2023arXiv230203540K} has been making waves\footnote{Pun intended.}. SPEAR-TTS has been proven to generally characterize anonymous speakers using a 3 second audio sample by solving text-to-semantic token and semantic token-to-acoustic token tasks separately. However, by doing this, the model becomes significantly slower than implementing a coupled version. It also requires some level of supervision, which is a significant disadvantage in relation to other TTS models. An even more recent model that is taking the field by storm is Google's SoundStorm \cite{2023arXiv230509636B}.  Unlike its predecessor AudioLM \cite{2022arXiv220903143B}, SoundStorm employs a new and far more efficient architecture that takes approximately 0.5 seconds to generate a 30-second segment of audio. SoundStorm is enabled to do so by the implementation of its bidirectional attention-based conformer \cite{2020arXiv200508100G} in conjunction with confidence-based parallel decoding. Furthermore, SoundStorm has experimented with the integration of the text-to-semantic modeling stage of SPEAR-TTS, and achieves audio faster despite its lower memory and processing needs.

One commonly observed design principle in SSL for speech recognition is the use of representation learning. However, its integration with self-supervised learning presents notable constraints. In order to address these limitations, a novel solution called BEST-RQ (BERT-based Speech pre-Training with Random-projection Quantizer, \cite{2022arXiv220201855C} has been proposed. BEST-RQ offers a simple yet effective self-supervised learning algorithm. The inclusion of a quantizer in this approach avoids the need for representation learning, thereby eliminating constraints on the model's architecture design. In particular, it outperforms previous approaches such as wav2vec 2.0 \cite{2020arXiv200611477B} and w2v-BERT \cite{2021arXiv210806209C} in certain tasks. With these impressive advances, it certainly becomes interesting as to how these methods could apply to TTS systems. However, BEST-RQ is tailored specifically for speech recognition tasks.

The persisting disadvantage that all of the aforementioned models face is that they are not task agnostic despite another specific advantage to SSL being that it helps models perform a wider breadth of downstream tasks. The pre-training phase using SSL allows the models to capture general knowledge and extract useful features from various domains without being task specific. Consequently, when fine-tuned on specific tasks with labeled data, the models can leverage the learned representations to quickly adapt and perform well across a range of tasks. SSL thus also expands the capabilities of models by providing them with a broader understanding of the data, enabling them to generalize better and transfer knowledge from one task to another. It effectively enhances the models' ability to handle diverse tasks by leveraging the latent information present in the unlabeled data during the initial learning phase. Despite this promising conjecture, beginning de novo to train purpose-built models with task-specific datasets has predominantly remained conventional procedure. 

Inspired by the achievements of SSL on language tasks, and motivated by the absence of an impressive universal pre-trained model for end-to-end speech processing tasks, WavLM was developed as a ``Large-Scale Self-Supervised Pre-Training" solution for ``Full-Stack Speech Processing" \cite{2022ISTSP..16.1505C}. WavLM is achieving a new state-of-the-art without the need for task-specific datasets during the training process. In its pre-training process, WavLM adopts a unique approach by simultaneously learning masked speech prediction and denoising. This method ensures that WavLM not only retains its ability to model speech content through masked speech prediction, but also enhances its potential for non-ASR tasks through speech denoising. Moreover, WavLM incorporates a gated relative position bias into the Transformer structure, which enhances its ability to accurately capture the sequential order of input speech.

Therefore, we propose experimenting with a combination of the BEST-RQ framework and the WavLM encoder. The simplistic yet effective framework of the BEST-RQ should benefit from a more task-agnostic encoder component and thus allow for the creation of a TTS system with prosodic, highly tunable speech. Specifically, in this work we compare the base setting of WavLM with a trained BEST-RQ model using the full 960 hours of LibriSpeech data \cite{inproceedings}. Similarly to STT, TTS has an array of possible use cases. In the context of continued research and aggressive growth, a TTS system can be used to generate synthetic training data for use in training ASR models. This would open up the ability to do very targeted training, or training on data that is very hard to obtain in large quantities, such as customer products, individual problem words, medical information, addresses, and names. Therefore, the development of a strong TTS system is very clearly the next step in advancing NLP as a whole. 

In Section \cref{sec:Background}, we will provide more detail on the main models WavLM (\cref{sec:WavLM}) and BEST-RQ (\cref{sec:BEST-RQ}), and provide more specific motivation for their combination in \cref{sec:The Combination}. In Section \cref{sec:Model Integration}, we will describe the process of the model merge, outline the main changes that need to be made to ensure intermodel compatibility, and provide the model parameters (\cref{sec: pretrain} and \cref{sec: training}). This leads us into a discussion on how the pretraining data was prepared (Section \cref{sec:Pre-Training Data Preparation}), and how we used SUPERB to benchmark our results (Section \cref{sec:Benchmarking TTS-Specific Downstream Tasks}). Finally, we detail our results and compare them to the published WavLM Base, Base+, and Large values in Section \cref{sec:Results}. We describe our conclusions based on the SUPERB results, and discuss the implications and proposed next steps in Section \cref{sec:Discussion and Conclusions}.  

\section{Background}
\label{sec:Background}

\subsection{WavLM}
\label{sec:WavLM}
WavLM is a pre-trained SSL speech model that takes the raw waveforms of speech signals as input. WavLM is built based on the Hidden-Unit BERT (HuBERT) framework \cite{DBLP:journals/corr/abs-2106-07447}; its two main components are a convolutional feature encoder and a Transformer encoder. The two main innovations in the model architecture that distinguishes WavLM from HuBERT are that (1) the Transformer structure uses gated relative position bias, and (2) randomly utterances are mixed into the raw input audio before feeding it to the encoder. The advantages that result from these additions are improvements to recognition and speaker discrimination tasks, respectively.

There exist a few WavLM versions differentiated by the data on which they were pre-trained. WavLM Base is pre-trained on 960 hours of LibriSpeech data \cite{inproceedings}, whereas WavLM Base Plus and WavLM Large are pre-trained on 60,000 hours of Libri-Light \cite{2019arXiv191207875K}, 10,000 hours of GigaSpeech \cite{DBLP:journals/corr/abs-2106-06909}, and 24,000 hours of VoxPopuli \cite{wang-etal-2021-voxpopuli}. The goal of this work is to compare the performance of the base setting of WavLM with the combination of the BEST-RQ vector quantization framework with the WavLM encoder. If the results indicate improvement, further research in comparing the more heavily pre-trained WavLM Base Plus and WavLM Large is of interest and possibility as WavLM Large has been shown to achieve state-of-the-art performance.

\subsection{BEST-RQ}
\label{sec:BEST-RQ}
HuBERT accepts continuous spoken inputs and employs the classic k-means algorithm trained on Mel-frequency cepstral coefficient (MFCC) features to identify hidden units from clustered segments of audio. Similarly, BEST-RQ accepts continuous spoken inputs in the form of log Mel spectrograms. However, instead of using k-means clustering, BEST-RQ is equipped with a random projection quantizer. This quantizer projects acoustic features to a randomly initialized matrix and uses the nearest vector in a randomly initialized codebook as the target label. 

\cite{2022arXiv220201855C} point out that the field's recent combination of SSL and representation learning result in some important limitations on speech recognition model architecture. Namely, that a model optimized for speech representation cannot reasonably be expected to perform optimally on downstream tasks. However, the limitation on the architecture design of the model is lifted when making the aforementioned change to the hidden unit identification process; the random projection quantizer is separated from the model and does not require representation learning. This implies that BEST-RQ, unlike the models before it, is better suited to be efficient when using very large amounts of unsupervised audio data and subsequently performing a wide array of downstream speech tasks. 

Another improvement that the BEST-RQ method facilitates is the fact that the quality of the initial vocabulary generated from MFCC features can be quite poor and computationally expensive to improve with a k-means teacher \cite{2022ISTSP..16.1179M}. Non-negligible performance gains result from the use of a quantizer in place that doesn't require a teacher. 

\subsection{The Combination}
\label{sec:The Combination}
WavLM is currently the highest ranked model in literature, performing especially well on speaker verification, speaker identification, and speaker diarization tasks, and BEST-RQ has already produced state-of-the-art results on multilingual tasks and presents promising performance on a wider expanse of speech tasks. Therefore, our proposition involves integrating the two frameworks to assess the extent to which the remarkable performance of WavLM, coupled with the superior suitability of BEST-RQ for downstream tasks, yields favorable outcomes, particularly in relation to semantic tasks, thus facilitating the development of a text-to-speech model that is both highly efficient and exceptionally accurate.

A focal point of our investigation lies in semantic tasks, as they play a crucial role in producing natural and contextually appropriate speech vital to a TTS system. By exploring the synergistic potential of the combined framework's performance specifically in this domain, we aim to ascertain the extent to which the integration enhances semantic understanding and expression in the synthesized speech output. This pursuit is driven by the increasing demand for sophisticated speech synthesis systems that can seamlessly generate natural-sounding speech across multiple languages and contexts.

\section{Model Integration}
\label{sec:Model Integration}

Hugging Face provides the bare WavLM Model Transformer\footnote{\url{https://huggingface.co/docs/transformers/model_doc/wavlm} } constructed to take in 16kHz sampled speech audio waveforms and output raw hidden states without any specific head on top. The BEST-RQ open source GitHub repository\footnote{\url{https://github.com/HarunoriKawano/BEST-RQ}} has been constructed in a way that is flexible enough that the encoder can be swapped out fairly easily. However, BEST-RQ is designed to take 80-dimensional log Mel spectrograms as input, as opposed to raw audio waveforms. Thus, a few important modifications to the original architecture must be made to sew these two frameworks together. 

\begin{figure*}[t]
    \centering
    \includegraphics[width=\textwidth]{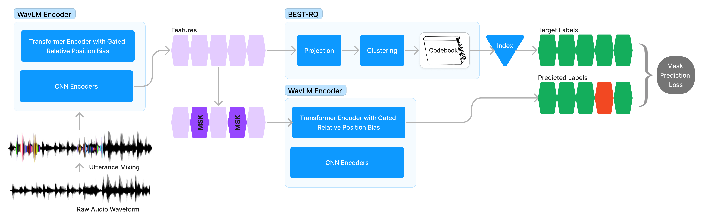}
    \caption{Combined Model Architecture}
    \label{fig:arch}
\end{figure*}

Firstly, the model must take 2 forward passes through WavLM. The raw audio samples are initially fed through the convolutional featurizer of the WavLM encoder. The outputs, the learned features, then take two distinct paths. Path 1 (top) in Figure \ref{fig:arch} outlines how the random projection quantizer projects the features to the randomly initialized matrix and finds a nearest vector in the randomly initialized codebook, which is used as the target label. Path 2 (bottom) follows the features that are then masked with the BEST-RQ technique and subsequently fed back to WavLM. Now, the predicted labels are generated by projecting the masked features and passing them into the Transformer encoder to obtain the last layer hidden states. This approach bears similarity to the masking strategy employed in Wav2Vec2 training, immediately indicating its potential effectiveness.

As an additional step, we have removed the two convolution layers at the bottom of the BEST-RQ model. This means that there are no further explicit reductions made in the temporal dimension to the input sequences. We have chosen to remove these as they were originally added for a separate model comparison with \cite{zhang2022pushing}. 

\subsection{Pre-training Parameters} \label{sec: pretrain}

During training, the probability of masking an input token is 0.01, and the duration of time the mask covers in an audio sequence is set to 0.4 seconds. A stride time of 0.01 seconds is used to define the time interval between consecutive masked segments. The size of the audio feature embeddings is 512. The size of the codebook used for quantization is set to 16, and the model employs 8192 of these codebooks. The encoder has 12 hidden Transformer layers, 768-dimensional hidden states, and 8 attention heads. We train with a batch size of 25 to match the WavLM Base training hyperparameters in \cite{2022ISTSP..16.1505C}. The model uses a linear warmup scheduler with 32k warm-up steps and 400k total training steps. During the warm-up steps, a linear warm-up scheduler is utilized to increase the learning rate from zero to a final value of 5e-4. The model utilizes the AdamW optimizer \cite{DBLP:journals/corr/abs-1711-05101} and a cross-entropy loss function for training.

Overall, the model has a total of 94.4 million trainable parameters and 6.4 million non-trainable parameters, resulting in a combined total of 100 million model parameters. The origin of the nontrainable parameters is the BEST-RQ codebook and projection matrix; they are separate from the rest of the model and do not learn. In total, the estimated model parameter size is 403.283 MB.

\subsection{Training Parameters} \label{sec: training}

We adopt the training settings introduced by the SUPERB project. To be specific, we use the same downstream models implemented in their work for each respective task and their exact datasets. In order to limit the search space for fine-tuning hyperparameters, we keep the pre-trained models frozen during the process. The downstream models are designed to leverage the weighted sum results of hidden states extracted from each layer of the pre-trained model. 

\section{Pre-Training Data Preparation}
\label{sec:Pre-Training Data Preparation}

In this work, we use the full 960 hours of LibriSpeech data \cite{inproceedings} to pretrain our model. Although already carefully curated, there are a few steps that we take to prepare the audio data before passing it as input. 

The first necessary step is to trim the audio to ensure both feasible and efficient input dimensions. In order to facilitate accurate result comparison, we aim to align with the published WavLM Base pre-training parameter values. Given that the average duration of a LibriSpeech audio file is approximately 14 seconds, and \cite{2022ISTSP..16.1505C} utilize a batch size of 350 seconds, we interpret this as accommodating 25 audio files per batch. To achieve this batch size on a single GPU, we randomly trim larger audio files to an arbitrary length of 14 times the LibriSpeech sample rate, which remains fixed at 16k. The audio files already shorter than this length are padded with zero-value samples to match the batch size. This ensures uniformity in the temporal dimension of the input data, allowing for a match of the parallel processing training efficiency.

Secondly, we follow Algorithm 1 as described by \cite{2022ISTSP..16.1505C}  to simulate overlapped speech on top of the original LibriSpeech data. Following the WavLM Base settings, we mix in only utterances (not noise) with a mixing probability of 0.20. These steps fully prep the LibriSpeech data to be appropriate input for the WavLM model.

\section{Benchmarking TTS-Specific Downstream Tasks}
\label{sec:Benchmarking TTS-Specific Downstream Tasks}

We train the model from the best pre-training checkpoint on several different audio datasets. Each time the model is trained for a specific downstream task, the training data used targets a semantic task useful to TTS applications. Specifically, we choose to focus on Spoken intent classification (IC), spoken slot filling (SF), and speech translation (ST). We benchmark all downstream tasks with the Speech processing Universal PERformance Benchmark (SUPERB) \cite{yang21c_interspeech} on the required datasets described below. 

In the context of testing NLP AI models, intent classification refers to the task of determining the underlying intention or purpose behind a given utterance. It involves classifying the input into predefined categories or classes that represent different intentions or desired actions. The objective is to accurately identify the specific goal or meaning expressed by the user's input. Intent classification plays a crucial role in natural language understanding and enables AI models to effectively comprehend and respond to user queries or commands. By accurately recognizing the intent, the system can provide appropriate and relevant responses or take the necessary actions. To perform intent classification, AI models are typically trained on labeled datasets where each input is associated with a specific intent class. These labeled examples serve as training data to teach the model the patterns and features indicative of each intent. The model then learns to generalize from the training data and make predictions on new, unseen inputs, assigning them to the most appropriate intent class. For this, we use the fluent speech commands dataset, which contains utterances used for controlling smart-home appliances or virtual assistants (whose job is to classify intent). 

Spoken slot filling refers to extracting specific information or data elements, known as slots, from audio inputs. The objective is to identify and fill these slots with the corresponding entities mentioned in the spoken utterances. The task of spoken slot filling is typically part of a larger dialogue or voice-based interaction system, where the AI model needs to understand and extract relevant information from the user's spoken input. Slots represent specific pieces of information that the system seeks to collect, such as dates, times, locations, names, or any other relevant entities based on the system's domain. To perform spoken slot filling, AI models are trained using annotated datasets where each spoken input is associated with labeled slot-value pairs. These pairs indicate the location of the slot within the spoken utterance and the corresponding value mentioned. The models learn to recognize patterns and linguistic cues to accurately identify the slots and extract the corresponding values. For this, we use the (un-pre-processed) Audio SNIPS dataset \cite{DBLP:journals/corr/abs-1805-10190}. 

Speech translation refers to the translation of spoken language from one source language to a target language in real time. The process typically involves two main steps: automatic speech recognition (ASR) and machine translation (MT). First, the ASR component converts the source language speech into its textual representation. Then, the MT component translates the recognized text into the target language. The task of speech translation poses several challenges due to the inherent complexities of spoken language, including variations in speech rate, accents, and background noise. Models used for speech translation are trained on large-scale multilingual datasets, incorporating both spoken audio and corresponding transcriptions in multiple languages. For this, we use CoVoST \cite{wang-etal-2020-covost}, a large-scale multilingual ST corpus based on Common Voice \cite{DBLP:journals/corr/abs-1912-06670}. 

\section{Results}
\label{sec:Results}

We now present the results of the SUPERB benchmark, which aims to assess the generalization capabilities of pre-trained language models on various downstream speech and language understanding tasks. The pre-trained models are obtained through self-supervised learning techniques on a diverse set of tasks and domains. Subsequently, they are fine-tuned on specific downstream tasks related to speech recognition and understanding.

For the Spoken Intent Classification (IC) task, we report test accuracies of 15.11 for the 100-hour training setting and 8.46 for the 960-hour setting. These results indicate that the pre-trained models struggled to generalize effectively to intent classification, with relatively low accuracies suggesting a lack of capturing the specific linguistic nuances and context required for accurate intent classification in the given datasets.

\begin{table}[htbp]
\centering
\label{tab:results}
\resizebox{150pt}{!}{%
\begin{tabular}{ccc}
\toprule
\textbf{Task} & \textbf{Metric} & \textbf{Test Value} \\
\midrule
IC & Test Accuracy (100h) & 15.11 \\
   & Test Accuracy (960h) & 8.46 \\
\midrule
\multirow{8}{*}{SF} & Test Loss & 1.4780 \\
& Test Slot\_type\_f1 & 43.25 \\
& Test Slot\_value\_cer & 82.68 \\
& Test Slot\_value\_wer & 92.52 \\
& Test Slot\_edit\_f1\_full & 23.2851 \\
& Test Slot\_edit\_f1\_part & 23.9784 \\
& Test WER & 63.44 \\
& Test CER & 43.70 \\
\midrule
KS & Test Accuracy & 25.41 \\
\hline
\end{tabular}
}
\caption{SUPERB Downstream Task Results}
\end{table}

\begin{table}[htbp]
\centering
\resizebox{\columnwidth}{!}{%
\begin{tabular}{|l|c|c||c|c|c|c|}
\hline
\multirow{3}{*}{Method} & \multirow{3}{*}{\# Params} &  \multirow{3}{*}{Corpus} & \multicolumn{3}{c|}{Semantics} & \multicolumn{1}{c|}{Content} \\
\cline{4-7}
 & & & \multicolumn{2}{c|}{IC} & SF & KS \\
\cline{4-7}
 & & & Acc $\uparrow$ & F1 $\uparrow$ & CER $\downarrow$  & Acc $\uparrow$ \\
\hline
WavLM Base &  94.70M &  LS 960 hr & 98.63 & 89.38 & 22.86 & 96.79 \\
\quad - w/o denoising task &  94.70M & LS 960 hr & 98.42 & 88.69 & 23.43 & 96.79 \\
\quad - w/o structure modification & 94.68M & LS 960 hr & 98.31 & 88.56 & 24.00 & 96.79 \\
WavLM Base+ &  94.70M & Mix 94k hr & 99.00 & 90.58 & 21.20 & 97.37 \\
WavLM Large & 316.62M & Mix 94k hr & 99.31 & 92.21 & 18.36 & 97.86 \\
Our model & 94.40M & LS 960 hr & 0.0846 & 0.4325 & 0.4370 & 0.2541 \\
\hline
\end{tabular}
}
\caption{SUPERB Results for WavLM models and the experimental combination model explored in this paper, for comparison.}
\label{tab:methods}
\end{table}

Regarding the Spoken Slot Filling (SF) task, we present multiple metrics, including test loss, slot type F1-score (43.25), slot value CER (82.68), slot value WER (92.52), slot edit F1-score (both full and partial) (23.2851 and 23.9784, respectively), and overall WER (63.44) and CER (43.70). The test loss of 1.4780 indicates the model's performance on the slot filling task, with lower values being more desirable. The slot type F1-score reflects moderate accuracy in identifying slot types, while the slot value CER and WER suggest challenges in accurately recognizing slot values within the spoken text. The slot edit F1-scores provide insight into the overall performance of the model in predicting correct slot values, and the reported WER and CER reveal suboptimal transcription performance. This places the model between the ranks of wav2vec and vq-wav2vec.

For the Keyword Spotting (KS) task, we observe a test accuracy of 25.41. The relatively low accuracy suggests that the pre-trained model did not perform well on this specific task. This places the model well below any of the benchmarked models. 

Overall, the results of the SUPERB benchmark comparisons indicate that the model explored in this paper did not optimally generalize to the downstream speech and language understanding tasks. The model severely underperformed in comparison to the state-of-the-art models also tested with the SUPERB benchmark, and the low accuracies and relatively high error rates suggest the need for further research. Although we had initially hoped for more competitive results, it is crucial to embrace and report on the aspects that did not work as expected. This investigation sheds light on areas that require further exploration and improvement. 

\section{Discussion and Conclusions}
\label{sec:Discussion and Conclusions}

Further research must be conducted to uncover the specific factors contributing to the less than desirable performance. Immediately, however, we recognize that a possible explanation involves the static nature of the random projection quantizer. The use of a random projection quantizer to determine target labels may have led to poor feature discrimination. The random projection quantizer is a static \emph{and random} component in the architecture, which poses certain challenges in terms of feature discrimination.

In speech recognition tasks, effective feature discrimination is essential for the model to distinguish between different acoustic patterns and speech segments. However, with a fixed random projection quantizer, the target labels assigned to the quantized acoustic features are initialized randomly- and most likely, therefore, poorly. Because they may be poorly predefined and do not change during training, as a result, the quantizer may not effectively capture the intricate and varying acoustic patterns present in the training data.

In contrast, dynamic or adaptive quantization techniques, such as k-means clustering or other learnable quantization methods, can adjust their target labels based on the input data distribution. This adaptability allows them to capture more nuanced patterns and fine-grained differences in acoustic features, potentially leading to better feature discrimination.

The static nature of the random projection quantizer in BEST-RQ may limit the model's ability to effectively represent and distinguish acoustic features, especially in complex speech recognition tasks or challenging environments with diverse acoustic variations. As a consequence, the model might struggle to achieve optimal performance, leading to less accurate recognition results.

To address this issue and improve feature discrimination, future research could explore the integration of more adaptive quantization techniques or investigate alternative methods to dynamically adjust the quantizer during training. By allowing the quantizer to adapt to the data distribution, the model can potentially learn more robust and discriminative representations of acoustic features, enhancing its overall performance in speech recognition tasks.

Another, related, possible reason for the observed performance stems from one of the previously mentioned key differences between how the WavLM and BEST-RQ models operate: spectrogram versus raw waveform input. 

In audio and speech processing, there is a significant distinction between extracting features directly from raw audio and obtaining features from log Mel spectrograms. Extracting features from raw audio involves working directly with the continuous audio waveform, capturing amplitude values or time-domain samples of the sound signal. This representation offers fine-grained temporal details but can be computationally intensive and challenging to process in various machine learning algorithms due to its high dimensionality. On the other hand, log mel spectrograms provide a widely used representation of audio signals. To create a mel spectrogram, the audio waveform is divided into short time frames, and a Fast Fourier Transform (FFT) is applied to each frame to convert it into the frequency domain. The resulting power spectrum is then transformed using a filterbank to create the Mel spectrogram, which mimics the human auditory system's frequency response. Taking the logarithm of the Mel spectrogram compresses the dynamic range and results in a more compact and robust representation suitable for machine learning models. Log Mel spectrograms are commonly used in automatic speech recognition (ASR), speaker recognition, and other audio-related tasks, effectively capturing both spectral and temporal information. The choice between raw audio and log Mel spectrograms usually depends on specific task requirements and the capabilities of the machine learning model being utilized. 

Because in this work we transitioned BEST-RQ's spectrogram approach to instead take in and quantize raw audio waveforms, we may have inadvertently decreased the quality of the extracted features. Initially, it seemed prudent to switch to using waveforms because they maintain phase information, a detail you lose during Fourier transformation. The fine-grained temporal details that are thereby retained can be essential for audio synthesis and waveform generation, given that these tasks require precise timing. However, some patterns in spectrograms occur at different time-frequency scales, which are perhaps more featurizable in terms of the random projection quantizer. 

It is a noted trend to architect deep learning models that mimic signal processing methods on both waveform and spectrogram inputs. Another possible avenue forward is to continue using raw audio processing, but implement a spectrogram-based loss function, which may give the best of both worlds.

In conclusion, this work reveals and analyzes the inherent incompatibility between the WavLM and BEST-RQ models. Despite these negative results, it is essential to recognize the significance of such investigations that highlight what does not work. Understanding the limitations and shortcomings of a particular approach is a critical step in the scientific process. By recognizing these gaps in our current understanding, we gain valuable insights that can pave the way for future advancements.

\appendix
\section{Appendix}\label{app: A}

\begin{table}[h]
\centering
\resizebox{\columnwidth}{!}{%
\begin{tabular}{l l}
\toprule
\textbf{Parameter} & \textbf{Value} \\
\midrule
Probability of masking an input token (mask\_prob) & 0.01 \\
Duration of time the mask covers in an audio sequence (mask\_time) & 0.4 seconds \\
Stride time between consecutive masked segments (stride\_time) & 0.01 seconds \\
Size of audio feature embeddings (input\_feature\_size) & 512 \\
Size of codebook for quantization (code\_book\_size) & 16 \\
Number of codebooks used (num\_code\_books) & 8192 \\
Number of temporal dimension reduction steps & 2 \\
Number of hidden Transformer layers in the encoder (encoder\_hidden\_size) & 12 \\
Dimension of hidden states & 768 \\
Number of attention heads & 8 \\
Batch size & 25 or 350s \\
Linear warm-up steps & 32k \\
Total training steps & 400k \\
Final learning rate & 5e-4 \\
Optimizer & AdamW \\
Loss function & Cross-entropy \\
\bottomrule
\end{tabular}
}
\caption{Pre-training parameters for the model.}
\label{tab:pretrain_parameters}
\end{table}

\begin{table}[h]
\centering
\begin{tabular}{l l}
\toprule
\textbf{Parameter} & \textbf{Value} \\
\midrule
Percentile & 100 \\
Mixing maximum length initialization & -1 \\
Mixing probability & 1 \\
Number of mixtures & 1 \\
Mixing with noise & True \\
Mixing noise probability & 0 \\
Number of noise mixtures & 1 \\
Utterance mixing probability & 0.2 \\
Utterance mixing & True \\
\bottomrule
\end{tabular}
\caption{Collator parameters for the model.}
\label{tab:collator_parameters}
\end{table}

\bibliographystyle{IEEEbib}
\bibliography{bib}

\begin{thebibliography}{10}

\bibitem{zhang2023speak}
Ziqiang Zhang, Long Zhou, Chengyi Wang, Sanyuan Chen, Yu~Wu, Shujie Liu, Zhuo Chen, Yanqing Liu, Huaming Wang, Jinyu Li, Lei He, Sheng Zhao, and Furu Wei,
\newblock ``Speak foreign languages with your own voice: Cross-lingual neural codec language modeling,'' arXiv, March 2023.

\bibitem{wang2023neural}
Chengyi Wang, Sanyuan Chen, Yu~Wu, Ziqiang Zhang, Long Zhou, Shujie Liu, Zhuo Chen, Yanqing Liu, Huaming Wang, Jinyu Li, Lei He, Sheng Zhao, and Furu Wei,
\newblock ``Neural codec language models are zero-shot text to speech synthesizers,'' arXiv, January 2023.

\bibitem{2023arXiv230203540K}
Eugene {Kharitonov}, Damien {Vincent}, Zal{\'a}n {Borsos}, Rapha{\"e}l {Marinier}, Sertan {Girgin}, Olivier {Pietquin}, Matt {Sharifi}, Marco {Tagliasacchi}, and Neil {Zeghidour},
\newblock ``{Speak, Read and Prompt: High-Fidelity Text-to-Speech with Minimal Supervision},''
\newblock {\em arXiv e-prints}, p. arXiv:2302.03540, Feb. 2023.

\bibitem{2023arXiv230509636B}
Zal{\'a}n {Borsos}, Matt {Sharifi}, Damien {Vincent}, Eugene {Kharitonov}, Neil {Zeghidour}, and Marco {Tagliasacchi},
\newblock ``{SoundStorm: Efficient Parallel Audio Generation},''
\newblock {\em arXiv e-prints}, p. arXiv:2305.09636, May 2023.

\bibitem{2022arXiv220903143B}
Zal{\'a}n {Borsos}, Rapha{\"e}l {Marinier}, Damien {Vincent}, Eugene {Kharitonov}, Olivier {Pietquin}, Matt {Sharifi}, Olivier {Teboul}, David {Grangier}, Marco {Tagliasacchi}, and Neil {Zeghidour},
\newblock ``{AudioLM: a Language Modeling Approach to Audio Generation},''
\newblock {\em arXiv e-prints}, p. arXiv:2209.03143, Sept. 2022.

\bibitem{2020arXiv200508100G}
Anmol {Gulati}, James {Qin}, Chung-Cheng {Chiu}, Niki {Parmar}, Yu~{Zhang}, Jiahui {Yu}, Wei {Han}, Shibo {Wang}, Zhengdong {Zhang}, Yonghui {Wu}, and Ruoming {Pang},
\newblock ``{Conformer: Convolution-augmented Transformer for Speech Recognition},''
\newblock {\em arXiv e-prints}, p. arXiv:2005.08100, May 2020.

\bibitem{2022arXiv220201855C}
Chung-Cheng {Chiu}, James {Qin}, Yu~{Zhang}, Jiahui {Yu}, and Yonghui {Wu},
\newblock ``{Self-supervised Learning with Random-projection Quantizer for Speech Recognition},''
\newblock {\em arXiv e-prints}, p. arXiv:2202.01855, Feb. 2022.

\bibitem{2020arXiv200611477B}
Alexei {Baevski}, Henry {Zhou}, Abdelrahman {Mohamed}, and Michael {Auli},
\newblock ``{wav2vec 2.0: A Framework for Self-Supervised Learning of Speech Representations},''
\newblock {\em arXiv e-prints}, p. arXiv:2006.11477, June 2020.

\bibitem{2021arXiv210806209C}
Yu-An {Chung}, Yu~{Zhang}, Wei {Han}, Chung-Cheng {Chiu}, James {Qin}, Ruoming {Pang}, and Yonghui {Wu},
\newblock ``{W2v-BERT: Combining Contrastive Learning and Masked Language Modeling for Self-Supervised Speech Pre-Training},''
\newblock {\em arXiv e-prints}, p. arXiv:2108.06209, Aug. 2021.

\bibitem{2022ISTSP..16.1505C}
Sanyuan {Chen}, Chengyi {Wang}, Zhengyang {Chen}, Yu~{Wu}, Shujie {Liu}, Zhuo {Chen}, Jinyu {Li}, Naoyuki {Kanda}, Takuya {Yoshioka}, Xiong {Xiao}, Jian {Wu}, Long {Zhou}, Shuo {Ren}, Yanmin {Qian}, Yao {Qian}, Jian {Wu}, Michael {Zeng}, Xiangzhan {Yu}, and Furu {Wei},
\newblock ``{WavLM: Large-Scale Self-Supervised Pre-Training for Full Stack Speech Processing},''
\newblock {\em IEEE Journal of Selected Topics in Signal Processing}, vol. 16, no. 6, pp. 1505--1518, Oct. 2022.

\bibitem{inproceedings}
Vassil Panayotov, Guoguo Chen, Daniel Povey, and Sanjeev Khudanpur,
\newblock ``Librispeech: An asr corpus based on public domain audio books,''
\newblock 04 2015, pp. 5206--5210.

\bibitem{DBLP:journals/corr/abs-2106-07447}
Wei{-}Ning Hsu, Benjamin Bolte, Yao{-}Hung~Hubert Tsai, Kushal Lakhotia, Ruslan Salakhutdinov, and Abdelrahman Mohamed,
\newblock ``Hubert: Self-supervised speech representation learning by masked prediction of hidden units,''
\newblock {\em CoRR}, vol. abs/2106.07447, 2021.

\bibitem{2019arXiv191207875K}
Jacob {Kahn}, Morgane {Rivi{\`e}re}, Weiyi {Zheng}, Evgeny {Kharitonov}, Qiantong {Xu}, Pierre-Emmanuel {Mazar{\'e}}, Julien {Karadayi}, Vitaliy {Liptchinsky}, Ronan {Collobert}, Christian {Fuegen}, Tatiana {Likhomanenko}, Gabriel {Synnaeve}, Armand {Joulin}, Abdelrahman {Mohamed}, and Emmanuel {Dupoux},
\newblock ``{Libri-Light: A Benchmark for ASR with Limited or No Supervision},''
\newblock {\em arXiv e-prints}, p. arXiv:1912.07875, Dec. 2019.

\bibitem{DBLP:journals/corr/abs-2106-06909}
Guoguo Chen, Shuzhou Chai, Guanbo Wang, Jiayu Du, Wei{-}Qiang Zhang, Chao Weng, Dan Su, Daniel Povey, Jan Trmal, Junbo Zhang, Mingjie Jin, Sanjeev Khudanpur, Shinji Watanabe, Shuaijiang Zhao, Wei Zou, Xiangang Li, Xuchen Yao, Yongqing Wang, Yujun Wang, Zhao You, and Zhiyong Yan,
\newblock ``Gigaspeech: An evolving, multi-domain {ASR} corpus with 10, 000 hours of transcribed audio,''
\newblock {\em CoRR}, vol. abs/2106.06909, 2021.

\bibitem{wang-etal-2021-voxpopuli}
Changhan Wang, Morgane Riviere, Ann Lee, Anne Wu, Chaitanya Talnikar, Daniel Haziza, Mary Williamson, Juan Pino, and Emmanuel Dupoux,
\newblock ``{V}ox{P}opuli: A large-scale multilingual speech corpus for representation learning, semi-supervised learning and interpretation,''
\newblock in {\em Proceedings of the 59th Annual Meeting of the Association for Computational Linguistics and the 11th International Joint Conference on Natural Language Processing (Volume 1: Long Papers)}, Online, Aug. 2021, pp. 993--1003, Association for Computational Linguistics.

\bibitem{2022ISTSP..16.1179M}
Abdelrahman {Mohamed}, Hung-yi {Lee}, Lasse {Borgholt}, Jakob~D. {Havtorn}, Joakim {Edin}, Christian {Igel}, Katrin {Kirchhoff}, Shang-Wen {Li}, Karen {Livescu}, Lars {Maaloe}, Tara~N. {Sainath}, and Shinji {Watanabe},
\newblock ``{Self-Supervised Speech Representation Learning: A Review},''
\newblock {\em IEEE Journal of Selected Topics in Signal Processing}, vol. 16, no. 6, pp. 1179--1210, Oct. 2022.

\bibitem{zhang2022pushing}
Yu~Zhang, James Qin, Daniel~S. Park, Wei Han, Chung-Cheng Chiu, Ruoming Pang, Quoc~V. Le, and Yonghui Wu,
\newblock ``Pushing the limits of semi-supervised learning for automatic speech recognition,'' 2022.

\bibitem{DBLP:journals/corr/abs-1711-05101}
Ilya Loshchilov and Frank Hutter,
\newblock ``Fixing weight decay regularization in adam,''
\newblock {\em CoRR}, vol. abs/1711.05101, 2017.

\bibitem{yang21c_interspeech}
Shu wen Yang, Po-Han Chi, Yung-Sung Chuang, Cheng-I~Jeff Lai, Kushal Lakhotia, Yist~Y. Lin, Andy~T. Liu, Jiatong Shi, Xuankai Chang, Guan-Ting Lin, Tzu-Hsien Huang, Wei-Cheng Tseng, Ko~tik Lee, Da-Rong Liu, Zili Huang, Shuyan Dong, Shang-Wen Li, Shinji Watanabe, Abdelrahman Mohamed, and Hung yi~Lee,
\newblock ``{SUPERB: Speech Processing Universal PERformance Benchmark},''
\newblock in {\em Proc. Interspeech 2021}, 2021, pp. 1194--1198.

\bibitem{DBLP:journals/corr/abs-1805-10190}
Alice Coucke, Alaa Saade, Adrien Ball, Th{\'{e}}odore Bluche, Alexandre Caulier, David Leroy, Cl{\'{e}}ment Doumouro, Thibault Gisselbrecht, Francesco Caltagirone, Thibaut Lavril, Ma{\"{e}}l Primet, and Joseph Dureau,
\newblock ``Snips voice platform: an embedded spoken language understanding system for private-by-design voice interfaces,''
\newblock {\em CoRR}, vol. abs/1805.10190, 2018.

\bibitem{wang-etal-2020-covost}
Changhan Wang, Juan Pino, Anne Wu, and Jiatao Gu,
\newblock ``{C}o{V}o{ST}: A diverse multilingual speech-to-text translation corpus,''
\newblock in {\em Proceedings of The 12th Language Resources and Evaluation Conference}, Marseille, France, May 2020, pp. 4197--4203, European Language Resources Association.

\bibitem{DBLP:journals/corr/abs-1912-06670}
Rosana Ardila, Megan Branson, Kelly Davis, Michael Henretty, Michael Kohler, Josh Meyer, Reuben Morais, Lindsay Saunders, Francis~M. Tyers, and Gregor Weber,
\newblock ``Common voice: {A} massively-multilingual speech corpus,''
\newblock {\em CoRR}, vol. abs/1912.06670, 2019.

\end{thebibliography}

\end{document}